\documentclass[preprint,superscriptaddress,prl,amsmath,aps]{revtex4-1}

\usepackage{graphicx}
\usepackage{amsmath}
\usepackage{array}
\usepackage[mathlines]{lineno}
\makeatletter
\newcommand*{\rom}[1]{\expandafter\@slowromancap\romannumeral #1@}
\makeatother
\begin{document}

    \title{Long-range non-Coulombic electron-electron interactions between LaAlO$_3$/SrTiO$_3$ nanowires}
	
	\author{Yuhe Tang}
    \affiliation{Department of Physics and Astronomy, University of Pittsburgh, Pittsburgh, Pennsylvania 15260, USA}
    \affiliation{Pittsburgh Quantum Institute, Pittsburgh, Pennsylvania 15260, USA}
    
    \author{Anthony Tylan-Tyler}
    \affiliation{Department of Physics and Astronomy, University of Pittsburgh, Pittsburgh, Pennsylvania 15260, USA}
    \affiliation{Pittsburgh Quantum Institute, Pittsburgh, Pennsylvania 15260, USA}
    
    \author{Hyungwoo Lee}
    \affiliation{Department of Materials Science and Engineering, University of Wisconsin-Madison, Madison, Wisconsin 53706, USA}
    
        \author{Jung-Woo Lee}
    \affiliation{Department of Materials Science and Engineering, University of Wisconsin-Madison, Madison, Wisconsin 53706, USA}
	
    \author{Michelle Tomczyk}
    \affiliation{Department of Physics and Astronomy, University of Pittsburgh, Pittsburgh, Pennsylvania 15260, USA}
    \affiliation{Pittsburgh Quantum Institute, Pittsburgh, Pennsylvania 15260, USA}
	
    \author{Mengchen Huang}
    \affiliation{Department of Physics and Astronomy, University of Pittsburgh, Pittsburgh, Pennsylvania 15260, USA}
    \affiliation{Pittsburgh Quantum Institute, Pittsburgh, Pennsylvania 15260, USA}
	
    \author{Chang-Beom Eom}
    \affiliation{Department of Materials Science and Engineering, University of Wisconsin-Madison, Madison, Wisconsin 53706, USA}
	
    \author{Patrick Irvin}
    \affiliation{Department of Physics and Astronomy, University of Pittsburgh, Pittsburgh, Pennsylvania 15260, USA}
    \affiliation{Pittsburgh Quantum Institute, Pittsburgh, Pennsylvania 15260, USA}
	
    \author{Jeremy Levy}
    \affiliation{Department of Physics and Astronomy, University of Pittsburgh, Pittsburgh, Pennsylvania 15260, USA}
    \affiliation{Pittsburgh Quantum Institute, Pittsburgh, Pennsylvania 15260, USA}
    \email[Corresponding author:]{jlevy@pitt.edu}
	
	\begin{abstract}

	The LaAlO$_3$/SrTiO$_3$  system exhibits unusual magnetic and superconducting behavior arising from electron-electron interactions whose physical origin is not well understood.
 Quantum transport techniques, especially those involving mesoscopic geometries, can offer insight into these interactions.  Here we report evidence for long-range electron-electron interactions in LaAlO$_3$/SrTiO$_3$ nanowires, measured through the phenomenon of frictional drag, in which current passing through one nanowire induces a voltage across a nearby electrically isolated nanowire.  
Frictional drag mediated by the Coulomb interaction is predicted to decay exponentially with interwire separation, but with the LaAlO$_3$/SrTiO$_3$ nanowire system it is found to be nearly independent of separation. 
Frictional drag experiments performed with three parallel wires demonstrates long-range frictional coupling even in the presence of an electrically grounded central wire.  
    Collectively, these results provides evidence for a new long-range non-Coulombic electron-electron interaction unlike anything previously reported for semiconducting systems.

	\end{abstract}
		
	\maketitle
   
	The heterointerface between the complex oxides LaAlO$_3$ and SrTiO$_3$ (LAO/STO) \cite{Ohtomo2004} exhibits a rich variety of electronically-tunable properties such as  superconductivity \cite{Caviglia2007,Caviglia2008},  magnetism \cite{brinkman2007,pai2016}, and spin-orbit coupling \cite{Shalom2010,Caviglia2010}. Many of these properties have been associated with strong gate-tunable electron-electron interactions \cite{Cheng2015,Cheng2016} which can be challenging to dissect using conventional transport methods.  The LAO/STO interface also exhibits a hysteretic metal-insulator transition \cite{Thiel2006}, which can be controlled locally using conductive atomic force microscopy (c-AFM) lithography \cite{ChengCen2008,ChengCen2009} and used to create a range of mesoscopic devices \cite{Sulpizio2014}.
    
    The transport technique of Coulomb drag \cite{Narozhny2016} (or more generally ``frictional drag") can provide unique insight into electron-electron interactions in the LAO/STO system.  When two electrical conductors are situated in close proximity, current driven through one (the ``drive") conductor may induce a voltage (or current) in the second (``drag") conductor. This effect was first proposed by Pogrebinskii \cite{Progrebinskii1977} as a method to probe correlations among the charge carriers of the system. Frictional drag measurements have been carried out in coupled 2D-3D semiconductor systems \cite{Solomon1989,Solomon1991}, coupled semiconductor 2DEGs \cite{Gramila1991, Gramila1992, Gramila1994, Solomon1991, Eisenstein1992} and graphene \cite{Li2016,Lee2016} systems, 1D-1D nanowires defined from semiconductor 2DEGs \cite{Debray2001,Tokura2006,Laroche2014}, and in coupled semiconductor quantum dots  \cite{Keller2016}. In these systems, the physical mechanism underlying frictional drag is dominated by Coulomb interactions. At large separations, non-Coulombic corrections can become apparent in some semiconductor devices \cite{Gramila1993,Tso1992,Tso1994}.

\begin{figure}
		\includegraphics[width=3.4in]{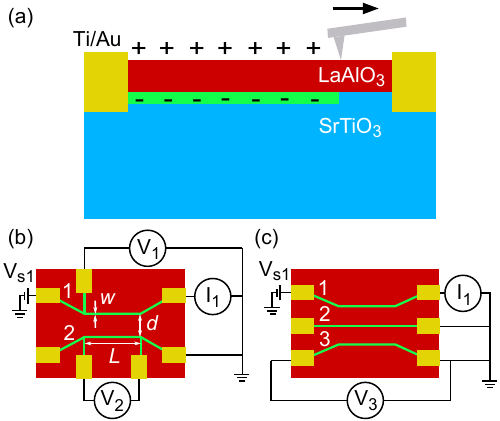}
		\caption{
			\label{Experiment}
Experimental setup. (a) Side-view of the nanowire fabrication process. A nanowire is created at the LAO/STO interface between two Ti/Au electrical contacts with c-AFM lithography. Protons ($+$) patterned on the surface by the AFM tip attract electrons ($-$) to the interface forming a nanowire (green area). (b) Top-view schematic of the double nanowire device with length $L$, width $w$, and wire separation $d$. The setup measures the induced drag voltage $V_\mathrm{drag}=V_2$ across wire 2 created by current $I_1$, which is induced by application of a voltage $V_{s1}$ across wire 1. (c) Schematic of a triple nanowire device where drag voltage $V_3$ induced by $I_1$ is measured. All three wires are grounded during the measurement.
				}
\end{figure}

Here we report frictional drag measurements performed on LAO/STO nanowire-based devices.  The device fabrication process is illustrated in Fig. 1(a). C-AFM lithography is used to define nanowires at the interface between 3.4 unit cells (uc) of LAO deposited on an STO substrate by pulsed laser deposition (PLD). Details of the sample growth and fabrication of electrical contacts are described elsewhere \cite{WaterCycle, ChengCen2008}. Positive tip voltages applied on the LAO surface produce locally conductive regions at the LAO/STO interface. The mechanism for the writing process is attributed to surface protonation \cite{WaterCycle, Bi2010}. A typical frictional drag system (illustrated in Fig. \ref{Experiment}(b)) is composed of two parallel nanowires with a width $w\sim10$ nm, length $L$ ranging between 400 nm and 1.5 $\mu$m, and separation $d$ ranging between $40$ nm and 1.5 $\mu$m. Devices consisting of three parallel nanowires (shown in Fig. \ref{Experiment}(c)) are also investigated. Except where noted otherwise, all measurements are performed below $T=$100 mK.
	In both double-wire (Fig. \ref{Experiment}(b)) and triple-wire (Fig. \ref{Experiment}(c)) device geometries,  frictional drag measurements are performed by sourcing a current $I_j$ in nanowire $j$ and measuring an induced voltage $V_i$ in nanowire $i$. All nanowires are connected to the same ground during the measurement. The current $I_j$ is produced by applying a voltage $V_{Sj}=V_{dc}+V_{ac}\cos\omega t$ to one end of nanowire $j$; the resulting current $I_j(\omega)$ and induced voltage $V_j(\omega)$ at frequency $\omega$ are measured using a lock-in amplifier. The resistance may then be expressed as a matrix $R_{ij}=dV_i/dI_j=V_i(\omega)/I_j(\omega)$, which is generally a function of the DC drive current $I_j$ (as well as other parameters such as temperature $T$ and applied magnetic field $\vec{B}$). 
The off-diagonal terms then define the drag resistance $R_{ij}$, characterizing the mutual friction between electrons in the drive and drag nanowires.  In order to ensure that the drag resistances $R_{ij}$ are not influenced by electron tunneling between the two nanowires, all measurments are performed well below the measured inter-wire breakdown voltage of each device.

\begin{figure}
		\includegraphics[width=3.4in]{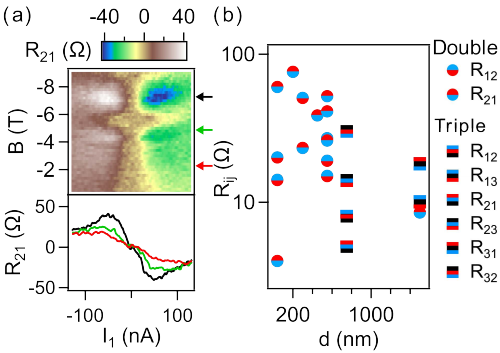}
		\caption{
			\label{Main data}
			Magnetic field and separation dependence of drag resistance (a) Top panel, drag resistance $R_{21}$ as a function of bias current $I_1$ and magnetic field $B$ (Device 2G). Bottom panel, line profiles of $R_{21}$ at $B$ = -7 T (black), -4.6 T (green) and -2 T (red). (b) Drag resistance $R_\textrm{ij}$ as a function of $d$ in the normal state regime in double- and triple-wire devices. Circle and square markers represent double and triple wire devices. Red, blue, and black represent drive, drag, and grounded wire, respectively. The relative position of three colors corresponds to the measurement configuration in the device.
		}
	\end{figure}

Typical results of a frictional drag measurement are shown in Fig. \ref{Main data}(a).  The nanowires are rendered non-superconducting by an applied magnetic field $\vec{B}=B \hat{z}$ (where $|B|>$0.2 T) applied perpendicular to the heterointerface. The magnitude of $R_{21}$ varies with $B$ and is antisymmetric in the drive current $I_1$. In frictional drag measurements, $R_{ij}$ is expected to be symmetric about $I_j=0$ as the interaction transfers momentum from the drive system to the drag system \cite{Debray2001, Tokura2006, Laroche2014, Levchenko2008}. Thus, when $I_j$ changes sign, so should $V_i$. The fact that $R_{ij}$ is antisymmetric with respect to $I_j$ indicates that inversion symmetry of the nanowires is broken somewhere and that quantum shot noise in the drive wire is primarily responsible for the drag voltage $V_2$ \cite{Levchenko2008}.
	
	In order to help identify the electron-electron interactions responsible for frictional drag in this system, we have created several devices with differing $L$ and $d$, as delineated in Table S\rom{1}  and S\rom{2} in Supplemental Material \cite{Supplement}. The maximum values for $|R_{ij}|$ for the magnetic field range explored (0.2 T$\leq B\leq$9 T) are plotted as a function of nanowire separation $d$ (Fig. \ref{Main data}(b)). Circle and square markers represent double-wire and triple-wire devices, respectively. Square markers are composed of red, blue and black segments that represent the arrangement of drive, drag and grounded wires, respectively. For example, a circle marker with blue on top corresponds to a measurement of $R_{12}$ in double wire device and a square one with color blue, black and red from top to bottom corresponds to a measurement $R_{13}$ for a triple-wire device. 
As shown in Fig. \ref{Main data}(b), the electron-electron interactions between the drive and drag nanowires exhibit large variations, with little if any explicit dependence on the nanowire separation $d$. This unusual scaling with distance is a significant departure from the expected behavior if the Coulomb interaction were responsible for the drag resistance. In the case when a Coulomb interaction gives rise to a drag voltage in coupled nanowires, Raichev et al \cite{Raichev2000} predict that $R_{ij}\propto e^{-4k_Fd}/\kappa^2$, where $k_F\sim$(10nm)$^{-1}$ is the Fermi wave vector, and $\kappa=4\pi\epsilon$ with $\epsilon>10000$ being the dielectric constant of STO \cite{Muller1979}. Such an exponential decay with distance is absent in our measurements. Moreover, the exceptionally large dielectric constant of STO should lead to a suppression of $R_{ij}$ by several orders of magnitude smaller compared with those measured in similar devices formed from other material systems \cite{Debray2001, Tokura2006, Laroche2014}; however, no such reduction is found. The weak scaling with separation, and the insensitivity to the large dielectric constant of STO indicate that the electron-electron interactions responsible for frictional drag are non-Coulombic in nature.
	
	\begin{figure}
		\includegraphics[width=3.4in]{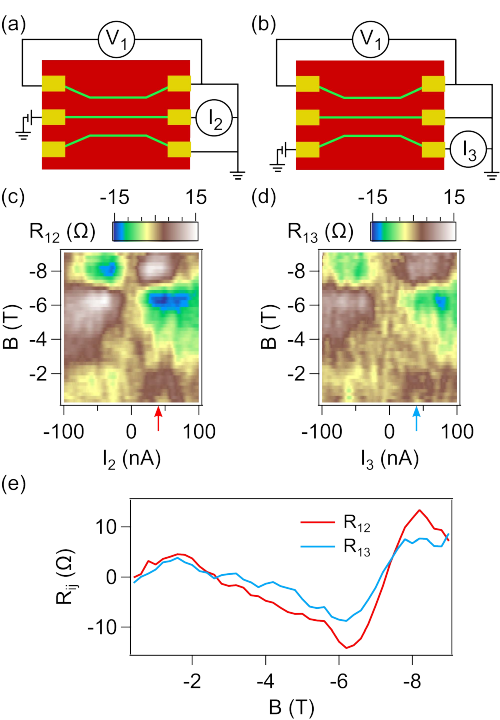}
		\caption{
			\label{Triple wire experiment}
			Triple wire experiment data. (a), (b) Schematics of triple-wire frictional drag. Drag voltage $V_\textrm{1}$ is measured from wire 1 with current sourced in wire 2 and 3 respectively. (c), (d) Drag resistance $R_{12}$ and $R_{13}$ corresponding to configurations in (a) and (b) plotted as a function of $B$. (e) Line profiles at $I_2$ and $I_3$ = 40 nA in (c) and (d)
		}
	\end{figure}

	In order to further explore the nature of the long-range interactions leading to frictional drag,  experiments with three parallel nanowires are investigated in detail. Schematics for two configurations (Figs. \ref{Triple wire experiment}(a, b)) yield measurements of $R_{12}$ and $R_{13}$, respectively (Figs. \ref{Triple wire experiment}(c,d)), as a function of drive current and magnetic field. 
    A comparison of $R_{12}$ and $R_{13}$ allows for the nanowire separation to be varied within a single device, and simultaneously probes the impact of introducing a central, grounded screening wire (for the case of $R_{13}$). 
Both the pattern as well as the magnitude of $R_{12}$ and $R_{13}$ are nearly identical, despite $d$ doubling (Fig. \ref{Triple wire experiment}(e)).  This result is consistent with the statistical findings summarized in Fig. \ref{Main data}(b). The frictional drag for the $R_{13}$ geometry is naively expected to be impacted by screening from the central wire. Instead, there is no discernable screening effect. The three-nanowire device geometry also enables one to ascribe the origin of the unique magnetic signature of the drag signal (i.e., Fig. \ref{Triple wire experiment}(c) and \ref{Main data} (a)) to properties of the drag wire and not the source wire.  

	\begin{figure}
		\includegraphics[width=3.4in]{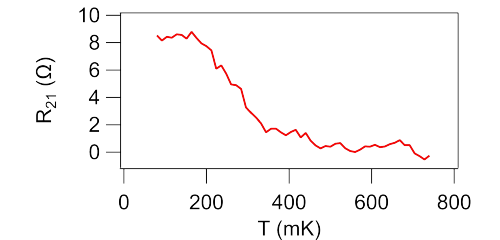}
		\caption{
			\label{T dependence}
			Temperature dependence of drag resistance $R_{21}$ between $T=80$ mK and $T=740$ mK at sourcing current $I_\textrm{2}=-100$ nA and $B=-9$ T (Device 2H). The drag resistance becomes negligible above $T=500$ mK for all of the devices investigated.
		}
	\end{figure}

	In 2D semiconductor drag systems, virtual phonon exchange was shown to be independent of distance \cite{Gramila1993, Tso1992, Tso1994}. The phonon-mediated coupling of the drag and drive systems, however, had a characteristice temperature scaling: the phonon mediated drag is expected to increase with increasing temperature \cite{Gramila1993,Narozhny2016}.  Fig. \ref{T dependence} shows the typical temperature dependence of frictional drag. The drag resistance decreases monotonically with increasing temperature,  becoming negligible  for $T>500$mK.  This temperature dependence is inconsistent with phonon-mediated frictional drag reported for 2D systems.
    
Frictional drag measurements between nanowires created on the LAO/STO heterointerface exhibit a strong, distance-insensitive coupling which indicate a non-Coulombic interaction.  The temperature dependence of this effect is incompatible with other known non-Coulombic interactions, such as virtual phonon exchange \cite{Gramila1993, Tso1992, Tso1994, Narozhny2016}. 
While these measurements do not specifically point to a particular coupling mechanism, there are candidates worth considering.   STO possesses a bulk cubic structure at room temperature that is unstable to an antiferroeistortive transition to a tetragonal phase below $T=$105 K \cite{Sakudo1971}. The ferroelastic domain structure gives rise to domain walls, which are correlated with anisotropic electronic phenomena observed at LAO/STO heterointerface \cite{Kalisky2013}. Ferroelastic domain walls are nominally insulating in the bulk, but they are also reported to be polar and mobile under applied electric fields.  The coupling of ferroelastic strain states and local surface potentials \cite{Honig2013} could potentially mediate long-range interactions through the insulating near-surface bulk STO layer.  Long-range couplings, whether mediated through ferroelastic domains or some other as-yet-unidentified mechanism, introduce a fascinating new element to the celebrated electronic properties of this oxide interface.

\pagebreak

\begin{center}
\textbf{\large Supplemental Material}
\end{center}
\onecolumngrid
\setcounter{table}{0}
\renewcommand{\thetable}{S\Roman{table}}
In the supplemental document, two tables are provided that list device parameters for double-wire and triple-wire devices. Parameters for double-wire devices include lengths and separations between nanowires, two-terminal, four-terminal resistances of each wire and drag resistances measured from each wire. In the table for triple-wire devices, each row corresponds to a different measurement configuration depending on which two nanowires are used for drive and drag wires. For example 3$A_{12}$ corresponds to the configuration where wire 1 and 2 are utilized and wire 3 is the grounded wire. $R_{12}$ is the drag resistance measured from wire 1 with wire 2 being the drive wire and vice-versa for $R_{21}$. Due to the limitation of the number of electrodes there is no four-terminal resistance in triple-wire device table.

\begin{table*}[h]
	\caption{
    \label{Distances}
    Double-wire device parameters.
	}
	\begin{ruledtabular}
		\begin{tabular}{ccccccccc}
			Device & $L$ (nm) & $d$ (nm) & $R_\textrm{2T, 1}$ (k$\Omega$) & $R_\textrm{2T, 2}$ (k$\Omega$)& $R_\textrm{11}$ (k$\Omega$) & $R_\textrm{22}$ (k$\Omega$) & $R_{12}$ ($\Omega$)& $R_{21}$ ($\Omega$)\\
			\hline
			2A & 400& 40 & 58-72 & 42-48 & 31.3-43.5 & 8.4-9.2 & 20 & 60 \\
			2B & 400 & 40 & 22-31 & 26-34 & \hspace{0.2cm}8.6-12.5 & 14.0-18.3 & 14 & 4 \\
			2C & 1000 & 300 & 37-47 & 25-46 & NA & \hspace{0.2cm}8.7-18.4 & 51 & 23 \\
			2D & 1000 & 450 & \hspace{0.2cm}40-108 & 29-34 & NA & \hspace{0.05cm}6.4-7.5 & NA & 39 \\
			2E & 1500 & 550 & 27-35 & 29-63 & NA & \hspace{0.2cm}7.8-11.8 & 15 & 52 \\
			2F & 1500 & 550 & 22-29 & 33-77 & NA & NA & 27 & 26 \\
			2G & 1500 & 550 & 23-36 & 22-51 & 11.5-16.0 & 3.7-5.5 & 19 & 41 \\
			2H & 1500 & 1500 & 17-27 & 22-37 & 10.2-14.8 & 2.7-4.1 & 10 & 9 \\
			2I & 400 & 200 & \hspace{0.2cm}50-127 & 27-37 & 29.7-97.8 & 5.2-6.7 & NA & 76 \\    
		\end{tabular}
	\end{ruledtabular}

\end{table*}

	\begin{table*}[h]
		\caption{
    \label{Distances1}
    Triple-wire device parameters. Subscripts $i$ and $j$ represent nanowires used in a configuration. $R_{ij}$ and $R_{ji}$ represent drag resistances measured from wire $i$ and $j$ respectively.
	}
	\begin{ruledtabular}
		\begin{tabular}{ccccccc}
			Config ($3\textrm{A}_{ij}$) & $L$ (nm) & $d$ (nm) & $R_{\textrm{2T}, i}$ (k$\Omega$) & $R_{\textrm{2T}, j}$ (k$\Omega$)&  $R_{ij}$ ($\Omega$) & $R_{ji}$ ($\Omega$)\\
			\hline
			3$\textrm{A}_\textrm{12}$ & 1500 & 750 & 26-41 & 26-34  & 14  & 5 \\
            3$\textrm{A}_\textrm{13}$ & 1500 & 1500 & 26-41  &  29-47 & 10  & 18 \\
            3$\textrm{A}_\textrm{23}$ & 1500 & 750 & 26-34 & 29-47 & 8  & 30 \\
            
		\end{tabular}
	\end{ruledtabular}

\end{table*}

\begin{thebibliography}{100}
\bibitem{Ohtomo2004}
A. Ohtomo and H. Y. Hwang, Nature \textbf{427}, 423 (2004).
\bibitem{Caviglia2007}
N. Reyren, S. Thiel, A. Caviglia, L. F. Kourkoutis, G. Hammerl, C. Richter, C. Schneider, T. Kopp, A.-S. Retschi, D. Jaccard, M. Gabay, D. A. Muller, J.-M. Triscone, and J. Mannhart, Science \textbf{317}, 1196 (2007).
\bibitem{Caviglia2008}
A. Caviglia, S. Gariglio, N. Reyren, D. Jaccard, T. Schneider, M. Gabay, S. Thiel, G. Hammerl, J. Mannhart, and J.-M. Triscone, Nature \textbf{456}, 624 (2008).
\bibitem{brinkman2007}
 A. Brinkman, M. Huijben, M. Van Zalk, J. Huijben, U.Zeitler, J. Maan, W. G. van der Wiel, G. Rijnders, D. H. Blank and H. Hilgenkamp, Nature Materials \textbf{6}, 493 (2007).
 \bibitem{pai2016}
Y.-Y. Pai, A. Tylan-Tyler, P. Irvin, and J. Levy, arXiv:1610.0789.
\bibitem{Shalom2010}
M. Ben Shalom, M. Sachs, D. Rakhmilevitch, A. Palevski, and Y. Dagan, Phys. Rev. Lett. \textbf{104}, 126802 (2010).
\bibitem{Caviglia2010}
A. D. Caviglia, M. Gabay, S. Gariglio, N. Reyren, C. Cancellieri, and J.-M. Triscone, Phys. Rev. Lett. \textbf{104}, 126803 (2010).
\bibitem{Cheng2015}
G. Cheng, M. Tomczyk, S. Lu, J. P. Veazey, M. Huang, P. Irvin, S. Ryu, H. Lee, C.-B. Eom, C. S. Hellberg, and J. Levy, Nature \textbf{521}, 196 (2015).
\bibitem{Cheng2016}
G. Cheng, M. Tomczyk, A. B. Tacla, H. Lee, S. Lu, J. P. Veazey, M. Huang, P. Irvin, S. Ryu, C.-B. Eom, A. Daley, D. Pekker, and J. Levy, Phys. Rev. X \textbf{6}, 041042 (2016).
\bibitem{Thiel2006}
S. Thiel, G. Hammerl, A. Schmehl, C. W. Schneider, and J. Mannhart, Science \textbf{313}, 1942 (2006).
\bibitem{ChengCen2008}
C. Cen, S. Thiel, G. Hammerl, C. W. Schneider, K. E. Andersen, C. S. Hellberg, J. Mannhart, and J. Levy, Nature Mater. \textbf{7}, 298 (2008).
\bibitem{ChengCen2009}
C. Cen, S. Thiel, J. Mannhart, and J. Levy, Science \textbf{323}, 1026 (2009).
\bibitem{Sulpizio2014}
J. A. Sulpizio, S. Ilani, P. Irvin, and J. Levy, Annual Review of Materials Research 44, 117 (2014).
\bibitem{Narozhny2016}
B. N. Narozhny and A. Levchenko, Rev. Mod. Phys. \textbf{88}, 025003 (2016).
\bibitem{Progrebinskii1977}
M. B. Pogrebinskii, Fiz. Tekh. Poluprovodn. \textbf{11}, 637 [Sov. Phys. Semicond. \textbf{11}, 372 (1977)].
\bibitem{Solomon1989}
P. M. Solomon, P. J. Price, D. J. Frank, and D. C. L. Tulipe, Phys. Rev. Lett. \textbf{63}, 2508 (1989).
\bibitem{Solomon1991}
P. M. Solomon, and B. Laikhtman, Superlattices Microstruct. \textbf{10}, 89 (1991).
\bibitem{Gramila1991}
T. J. Gramila, J. P. Eisenstein, A. H. MacDonald, L. N. Pfeiffer, and K. W. West, Phys. Rev. Lett. \textbf{66}, 1216 (1991).
\bibitem{Gramila1992}
T. J Gramila., J. P. Eisenstein, A. H. MacDonald, L. N. Pfeiffer, and K. W. West, Surf. Sci. \textbf{263}, 446 (1992).
\bibitem{Gramila1994}
T. J. Gramila, J. P. Eisenstein, A. H. MacDonald, L. N. Pfeiffer, and K. W. West, Physica (Amsterdam) \textbf{197B}, 442 (1994).
\bibitem{Eisenstein1992}
J. P. Eisenstein, Superlattices Microstruct. \textbf{12}, 107 (1992).
\bibitem{Li2016}
J. I. A. Li, T. Taniguchi, K. Watanabe, J. Hone, A. Levchenko, and C. R. Dean, Phys. Rev. Lett. \textbf{117}, 046802 (2016).
\bibitem{Lee2016}
K. Lee, J. Xue, D. C. Dillen, K. Watanabe, T. Taniguchi, and E. Tutuc, Phys. Rev. Lett. \textbf{117}, 046803 (2016).
\bibitem{Debray2001} 
P. Debray, V. Zverev, O. Raichev, R. Klesse, P. Vasilopoulos, and R. S. Newrock, J. of Phys.: Cond. Matt. \textbf{13}, 3389 (2001).
\bibitem{Tokura2006} 
M. Yamamoto, M. Stopa, Y. Tokura, Y. Hirayama, and S. Tarucha, Science, \textbf{313}, 204 (2006).
\bibitem{Laroche2014}
D. Laroche, G. Gervais, M. P. Lilly, and J. L. Reno, Science, \textbf{343}, 631, (2014).
\bibitem{Keller2016} 
A. J. Keller, J. S. Lim, D. S\'{a}chez, R. L\'{o}pez, S. Amasha, J. A. Katine, H. Shtrikman, and D. Goldhaber-Gordon, Phys. Rev. Lett. \textbf{117}, 066602 (2016).
\bibitem{Gramila1993} 
T. J. Gramila, J. P. Eisenstein, A. H. MacDonald, L. N. Pfeiffer, and K. W. West, Phys. Rev. B \textbf{47}, 12957 (1993).
\bibitem{Tso1992}
H. C. Tso, P. Vasilopoulos, and F. M. Peeters, Phys. Rev. Lett. \textbf{68}, 2516 (1992).
\bibitem{Tso1994}
H. C. Tso,P. Vasilopoulos, and F. M.Peeters, Surf. Sci. \textbf{305}, 400 (1994).
\bibitem{WaterCycle} 
 K. Brown, S. He, D. J. Eichelsdoerfer, M. Huang, I. Levy, H. Lee, S. Ryu, P. Irvin, J. Mendez-Arroyo, C.-B. Eom, C. A. Mirkin, and J. Levy, Nature Comm. $\bf{7}$, 10681 (2016).
 \bibitem{Bi2010} 
F. Bi, D. F. Bogorin, C. Cen, C. W. Bark, J.-W. Park, C.-B. Eom, and J.Levy, Appl. Phys. Lett. $\bf{97}$, 173110 (2010).
\bibitem{Levchenko2008}
A. Levchenko and A. Kamenev, Phys. Rev. Lett. \textbf{101}, 216806 (2008).
\bibitem{Supplement}
See Supplemental Material at [URL will be inserted by publisher] for parameters of double- and triple-wire devices
\bibitem{Raichev2000}
O. E. Raichev and P. Vailopoulos, Phys. Rev. B \textbf{61}, 7511 (2000).
\bibitem{Muller1979}
K. A. M\"{u}ller and H. Burkard, Phys. Rev. B \textbf{19}, 3593 (1979).
\bibitem{Sakudo1971}
T. Sakudo and H. Unoki, Physical Review Letters 26, 851 (1971).
\bibitem{Kalisky2013}
B. Kalisky, E. M. Spanton, H. Noad, J. R. Kirtley, K. C. Nowack, C. Bell, H. K. Sato, M. Hosoda, Y. Xie, Y. Hikita, et al, Nat. Mat. \textbf{12}, 1091 (2013).
\bibitem{Honig2013}
M. Honig, J. A. Sulpizio, J. Drori, A. Joshua, E. Zeldov, and S. Ilani, Nat. Mat. \textbf{12}, 1112 (2013).
\end{thebibliography}
\end{document}